\documentclass[useAMS,usenatbib]{mn2e}
\usepackage{journals}
\usepackage{times}

\newcommand{\sub}[1]{_{\mathrm{#1}}}

\newcommand{\ergscm}[1]{$10^{#1}$~erg~cm$^{-2}$~s$^{-1}$}
\newcommand{\ergcm}[1]{$10^{#1}$~erg~cm$^{-2}$}

\usepackage{url}
\usepackage{amssymb}
\usepackage{graphicx}

\title[Double-peaked type I bursts in the RB]{Double-peaked thermonuclear bursts at the soft-hard state transition in the Rapid Burster}
\author[Bagnoli et al.]{T. Bagnoli$^{1,2}$\thanks{E-mail:
t.bagnoli@sron.nl},
J.J.M. in 't Zand$^{1}$,
A. Patruno$^{2}$
and A.L. Watts$^{2}$
\\
$^{1}$SRON Netherlands Institute for Space Research,
Sorbonnelaan 2, 3584 CA Utrecht, The Netherlands\\
$^{2}$Astronomical Institute ``Anton Pannekoek'', University of Amsterdam,
Postbus 94249, 1090 GE Amsterdam, The Netherlands
\\
\\
\\
\textup{Accepted 2013 October 25. Received 2013 October 21; in original form 2013 August 19}
}

\begin{document}

\date{}

\pagerange{\pageref{firstpage}--\pageref{lastpage}} \pubyear{}

\maketitle

\label{firstpage}

\begin{abstract}

Long suspected to be due to unstable accretion events,
the type II bursts exhibited by the Rapid Burster (RB, or MXB 1730-335)
still lack an exhaustive explanation.
Apart from type II bursts,
the transient RB also shows 
the better-understood
thermonuclear shell flashes known as type I bursts.
In search of links between these two phenomena,
we carried out a comprehensive analysis
of all \textit{Rossi X-ray Timing Explorer}
observations of the RB
and found six atypical type I bursts,
featuring a double-peaked profile
that is not due to photospheric radius expansion.
The bursts appear in a phase of the outburst decay
close to the onset of the type II bursts,
when the source also switches from the high/soft to the low/hard state.
We also report the discovery of a simultaneous
low-frequency quasi-periodic oscillation
present in the persistent emission as well as in the burst decaying emission.
We discuss several scenarios to understand the nature
of the peculiar bursts and of the accompanying oscillation,
as well as their possible relation 
with each other and with the type II burst phenomenon.
We favour a model consisting of two accretion channels,
one polar and one equatorial, in a source viewed at low inclination.
We also, for the first time, clearly identify the atoll nature of the RB.

\end{abstract}


\section{Introduction}

The Rapid Burster \citep*[RB; e.g.][]{1993SSRv...62..223L}
is a transiently-accreting neutron star (NS) low-mass X-ray binary (LMXB),
with outbursts recurring in less than one year \citep{2002A&A...381L..45M}
and peak luminosities up to nearly half the Eddington limit \citep[][hereafter B13]{2013MNRAS.431.1947B}.
It is located in the globular cluster Liller 1, at a distance of $7.9 \pm 0.9$~kpc
\citep*{2010MNRAS.402.1729V}.
It is unique, in that it shows both
type I bursts, which are X-ray eruptions due to the heating and cooling of the NS photosphere
after a thermonuclear shell flash of accreted material (see e.g. reviews by
\citealt{2006csxs.book..113S}
and \citealt{2008ApJS..179..360G}), and
type II bursts, most likely due to the release of gravitational energy
during sudden accretion episodes \citep*{1978Natur.271..630H}.
While the former are relatively well understood and observed in over a hundred sources,
it remains unclear why the latter have been unambiguously detected in only two NS LMXBs,
the other one being the slowly rotating (2~Hz) Bursting Pulsar \citep{1996Natur.379..799K}.

The Proportional Counter Array (PCA) onboard the \textit{Rossi X-ray Timing Explorer} (\textit{RXTE})
observed 125 type I and about 7050 type II bursts from the RB.
A complete analysis (B13) of the type I burst sample
shows that type I bursts appear in the RB
over a remarkably large range of mass accretion rates,
recurring faster as its bolometric luminosity increases
to a large fraction of the Eddington limit,
a behaviour only shared with the 11~Hz pulsar IGR J17480-2446 \citep{2012ApJ...748...82L}
among over a hundred bursting LMXBs (see e.g. \citealt{2003A&A...405.1033C}).
The known NS spin frequencies in these systems
range between 245 and 620~Hz \citep{2012ARA&A..50..609W},
and the exceptional bursting behaviour of IGR J17480-2446 
can be explained only thanks to its slow NS spin \citep{2012ApJ...748...82L}.
This indicates the RB should be a slow rotator too (B13).

Type II bursts instead appear only after the outburst has significantly decayed,
usually below a bolometric persistent flux of
$F\sub{pers} \approx 3 \times$\ergscm{-9} \citep{1999MNRAS.307..179G}.
They often show a relaxation-oscillator behaviour,
meaning that the fluence in a type II burst
is proportional to the waiting time to the next one.
These properties  hint at disc-magnetosphere interactions
that are largely thought to control the type II burst phenomenon
\citep{1977ApJ...214..245B,1979ApJ...227..987B,1977ApJ...217..197L,1993ApJ...402..593S}.
Type II bursts in the RB are often accompanied
by low-frequency quasi periodic oscillations (LF QPOs) in the persistent emission,
both at frequencies $\nu_0$ between 2 and 4~Hz,
and at much slower values centred about an average of 42~mHz.
They both cease right before the onset of a type II burst,
so it is likely that they originate in the same mass reservoir whose spasmodic accretion
is responsible for the type II burst \citep{1988ApJ...324..379S,1992MNRAS.258..759L}.

Much theoretical work
\citep[e.g.][]{2008MNRAS.386..673K,2010MNRAS.406.1208D,2012MNRAS.421...63R,2012MNRAS.420..416D}
has focussed on the dynamical role of the magnetic field,
which can act as a ``gating'' mechanism to the accretion flow
once the mass accretion rate drops below a certain threshold.
These models 
could explain the episodic accretion
that is responsible for the type II burst phenomenon,
but also LF QPOs at low mass accretion rates, as those observed
in the lightcurves of the accreting millisecond X-ray pulsars (AMXPs)
SAX J1808.4-3658 \citep{2009ApJ...707.1296P}
and NGC 6440 X-2 \citep{2013ApJ...771...94P}.

We report here on six peculiar type I bursts 
that stand out among the RB PCA sample
because of a unique double-peaked structure
first noticed by \citet{2001MNRAS.321..776F}.
We show that four of these bursts appear at the transition
from the high/soft to the low/hard state \citep{2006csxs.book...39V},
and that this takes place at the time of the first appearance of type II bursts.
We exclude the possibility that their double peaks are due to photospheric radius expansion
(PRE; see e.g. \citealt{1993SSRv...62..223L}).
We also find a previously unreported QPO
at a frequency of about 0.25~Hz.
The QPO becomes visible in the persistent emission
a few days before the double-peaked bursts,
peaks in amplitude at the time the bursts appear,
even becoming directly visible in the lightcurve,
and fades out over the next few days.

We describe the spectral and timing properties
of the double-peaked type I bursts
and the LF QPO accompanying them,
and investigate whether they can shed further light
on the nature of the interaction between
accretion and magnetic-field effects in the enigmatic RB.
Section~\ref{sec:obsana} describes the dataset
and discusses the data analysis.
Section~\ref{sec:res} presents the results of our analysis,
and finally Section~\ref{sec:dis} discusses
the interpretation of these results
in the context of models and observations of other sources.

	\section{Observations}\label{sec:obsana}

Most measurements described here were obtained through observations with the PCA.
The PCA \citep{2006ApJS..163..401J} is a non-imaging system
consisting of five co-aligned proportional counter units (PCUs)
that combine to a total effective area of 6500~cm$^2$ at 6~keV, in a 2 to 60~keV bandpass.
The photon energy resolution is 18 per cent full width at half maximum (FWHM) at 6~keV
and the time resolution is programmable down to about 1~$\mu$s.
The field of view (FOV) is approximately circular,
with radius 1$\degr$ (full width to zero response).
The RXTE observations of the RB include a mix of data.
There are observations directly pointed at the source
that are contaminated by the presence of 4U 1728-34
at an angular distance of 0.56$\degr$ \citep{2008ApJS..179..360G},
and offset pointings that do not suffer from such contamination, albeit at the price
of a smaller effective area.
The total exposure time obtained on the RB while it was active is 0.91 Msec.
For further details on source confusion in the FOV, burst identification,
and discrimination between type I and type II bursts,
we refer the reader to B13.

Six type I bursts have a peculiar double-peaked time profile.
They are numbered 35, 36, 37, 58, 118 and 119,
following their chronological position
among the 125 type I burst in the PCA sample (B13).
Three of the six double-peaked type I bursts subject of this paper
occurred during one outburst which is shown in Fig.~\ref{fig:outburst}.
The burst lightcurves (Fig.~\ref{fig:humps})
were constructed employing 0.125-s resolution \textsc{Standard 1} data.
Typically, type I X-ray bursts of the RB show a two-component decay profile,
with a steeper 10-s long initial decay and a shallower 100-s long tail,
whose relative amplitude varies  (B13).
We constructed an average type I burst profile
employing 22 bursts which we chose
because they show the longest and brightest decays.
This average burst profile is plotted
against each burst in Fig.~\ref{fig:humps}.

We define a few variables to help us quantify
the morphology of the dip between the peaks.
First, we applied an exponential fit to the first 200~s of
the slow-decay portion of the burst after the second peak.
We extrapolated this fit backwards to determine the time $t\sub{dip,start}$
when it intercepts the first peak of the burst.
The dip duration $\Delta t\sub{dip}$ is the time between $t\sub{dip,start}$
and the time of the second peak.
The dip depth $D$ is the ratio of the number of photons
lacking under the extrapolated curve
during the interval $\Delta t\sub{dip}$
over the total number of observed photons between the burst start- and end times
(defined as in B13).

Time-resolved spectroscopy of the burst data
was performed following the same procedure and tools as in B13.
The particle and cosmic background were determined with \textsc{ftool pcabackest}
and subtracted from all spectra.
Response matrices were generated with \textsc{pcarsp} (v. 11.7.1).
All active PCUs were employed, and a correction was applied for
deadtime (although it is always small for the RB).
Following the \textit{RXTE} Cookbook prescription\footnote{See \url{http://heasarc.nasa.gov/docs/xte/recipes/cook_book.html}},
the analysis was limited to the $3-25$~keV range,
and a systematic error of 0.5\% was added
in quadrature to the statistical error of the flux in each channel.
The burst data were divided into a number of time bins,
varying in duration to keep the photon count,
and hence the relative error on the derived quantities, approximately constant.

\section{Results}\label{sec:res}

	\subsection{The outbursts}\label{sec:outb}

\begin{figure}
  	\includegraphics[width=0.95\columnwidth]{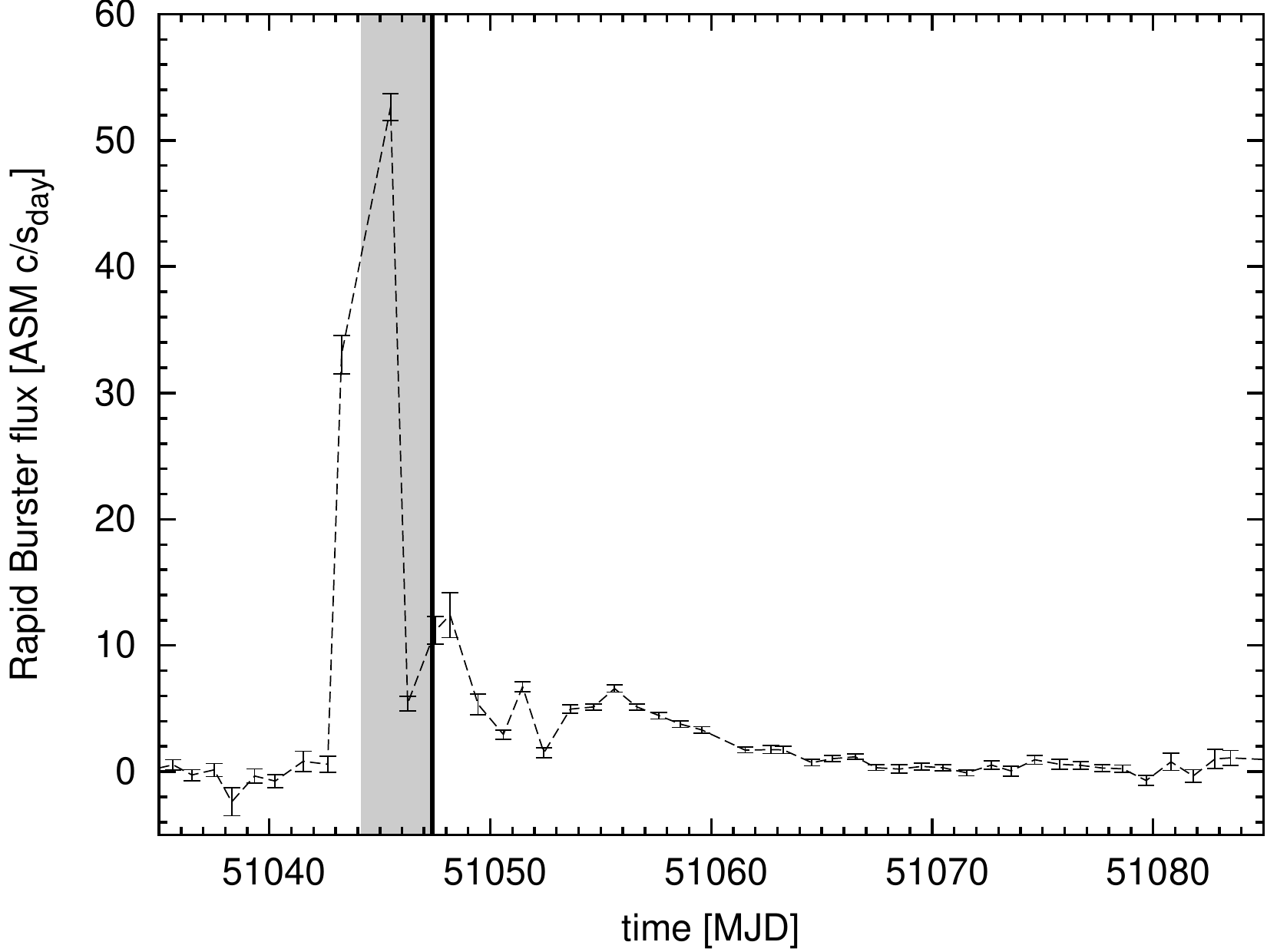}
	\caption{\small{Daily average 3-12~keV flux of the RB
measured with the All-Sky Monitor (ASM, \citealt{1996ApJ...469L..33L})
on board the \textit{RXTE}
during the August-September 1998 outburst.
The three double-peaked type I bursts 35 through 37
occur within $\lesssim$2~h (thick black line).
The 0.25~Hz QPO (see Sec.\ref{sec:QPO}) is visible in the power spectra
for about three days,
from the beginning of the PCA observations until the double-peaked bursts (grey-shaded area).
Long, flat-topped and very energetic type II bursts appear at the same time
(see Sec.~\ref{sec:outb}).
}}
	\label{fig:outburst}
\end{figure}

Typical RB outbursts observed by \textit{RXTE} prior to November 1999
tend to have a fast rise of 1 to 3 days, followed by an exponential decay
with an e-folding decay time of about 8 days.
Initially only type I bursts are observed.
Type II bursts activity starts about
18 days into the outburst,
when typically the bolometric flux decreased to a level of
$F\sub{pers} \approx 3 \times$\ergscm{-9}.
The first type II bursts to appear
(the so-called mode-0 type II bursts) are
long (200-300~s), bright (up to 3000~c/s/PCU), flat-topped
and short-lived, usually being visible for only a couple of days.
We inspected all the light curves visually and
only found 18 mode-0 type II bursts.
Their rarity is due both to the short outburst fraction in which they are produced
and to their long recurrence time,
which for RB type II bursts is proportional to the fluence in a burst
(the so-called relaxation-oscillator behaviour, \citealt{1993SSRv...62..223L}).
Shorter, more frequent (the so-called mode I and mode II) type II bursts
then follow for another couple of weeks
until the end of the outburst \citep{1999MNRAS.307..179G}.

The August-September 1998 outburst during which bursts 35 through 37
appeared has an atypical lightcurve shape (see Fig.\ref{fig:outburst}),
featuring a much faster decay,
as well as much earlier mode-0 type II bursts,
which could have started already during the outburst rise
\citep{1998ATel...32....1F}.
Although no PCA coverage is available for more than three days before,
the timing of the three double-peaked bursts seems also quite interesting.
As can be seen in Fig.\ref{fig:outburst}, these double-peaked bursts
occur during the initial phase of type II bursts,
at a time when the source is experiencing rapid and large variations in flux (``reflares'').
These considerations make us confident
that the peculiar sequence of double-peaked type I bursts
occurs close to the onset of the type II bursting activity.

Burst 58 was observed eight days after the start time
of the PCA observations of the October 1999 outburst.
In the time leading to it, twelve type I bursts are observed,
and no type II bursts.
No observation is available in the three days prior to burst 58,
and not again until four days later.
Short (mode-I) type II bursts are first observed ten days after burst 58.
It seems plausible that this could be a more typical outburst,
and that the short-lived phase of
strong, long and flat-topped (mode-0) type II bursts
was missed because of the rather limited coverage
around the time it could have occurred.

Bursts 118 and 119 took place during the last RB outburst observed by \textit{RXTE},
in January 2010.
Less than 12~ks of PCA data are available for this outburst, in 5 observations spanning only 3 days.
It is therefore difficult to assess to what extent
the outburst has already decayed when these bursts occur.
However, starting in 2000 all outbursts recurred within a shorter time
and peaked at significantly lower intensities (Masetti 2002).
Therefore, the initial type-I-only phase
of the outburst can be significantly shorter.
Judging from the ASM lightcurve,
bursts 118 and 119 occurred no longer than 2~days
after the beginning of the outburst.
None the less,
the presence of six more type I bursts,
the burst recurrence time (see Sec.~\ref{sec:lightcurves}),
the count rate in the persistent emission
and the lack of type II bursts
are all consistent with what is observed in the other four double-peaked bursts,
which indicates a similar accretion state.

	\subsection{The double-peaked bursts}\label{sec:morph}

		\subsubsection{Lightcurves}\label{sec:lightcurves}

\begin{figure*}
  	\includegraphics[width=1.95\columnwidth]{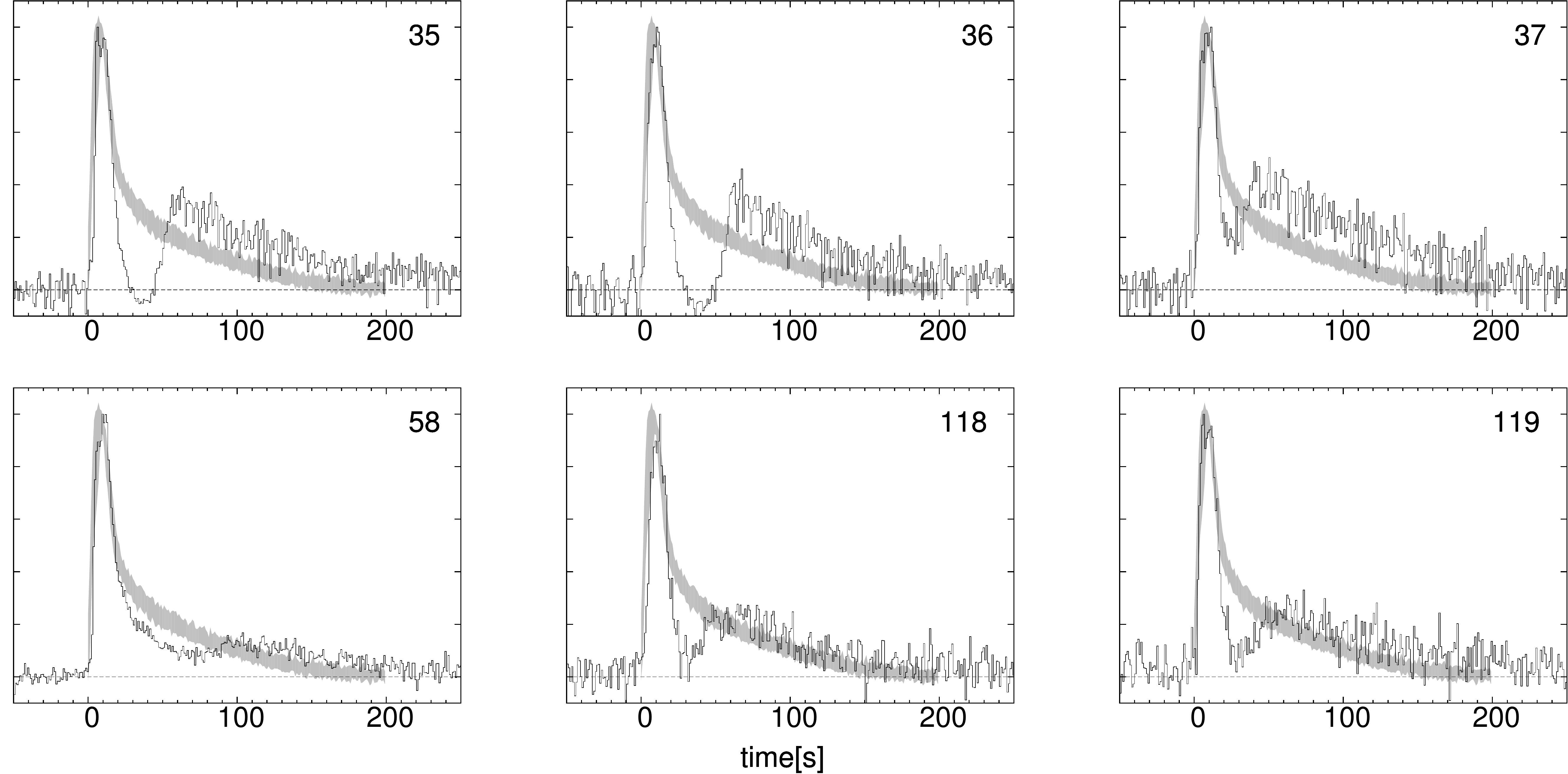}
	\caption{\small{The six bursts studied here.
The time axis has been shifted so that at the burst start time $t = 0$.
The subtracted background level is measured from a 500~s time interval prior to the burst.
The curves are normalized to their individual peak rate.
The zero level is indicated by a dashed line.
Each burst is plotted against the average profile of the 22 single-peaked type I bursts
featuring the longest and brightest decays (see Sec.~\ref{sec:morph}).
The shaded region indicates the standard deviation of the sample.
The large oscillations in the decays of bursts 35 through 37 are not due to random (white) noise,
but to the 0.25~Hz QPO.
They are clearly still visible in the shallower dip of burst 37,
but absent in the deeper dip of burst 35.
Burst 58 reaches a much larger count rate in its first peak,
which makes the second peak and the QPO harder to see in this scaling.}}
	\label{fig:humps}
\end{figure*}

Fig.~\ref{fig:humps} shows the six double-peaked type I bursts under analysis
and Table~\ref{tab:spectra} lists their characteristics.
Bursts 35 through 37 occurred with a recurrence time of about 2.2~ks,
which is in the middle of the range of observed values 
of all RB type I bursts (B13).
In bursts 35 and 36 the dip clearly lasts longer
($\Delta t\sub{dip} =46$~s)
and the flux even drops to about 5 per cent below the pre-burst emission level.
The dip in burst 37 is instead shorter, at 25~s,
and the flux only drops to about 20 per cent of the net burst peak count rate.
While the QPO is not visible during the dips of bursts 35 and 36,
it is still present in that of burst 37.

The second peak is about 40 to 50 per cent as large as the first one.
It is higher in burst 37, where it occurs earlier.
The QPO modulation is clearly visible again in the second portion of the decays.

Looking at Fig.~\ref{fig:humps},
it is clear that the post-dip portion of the decay
is significantly brighter than in the rest of the type I bursts sample.
As a result, the bursts are also longer.
If however one only looks at the minima of the QPO,
the decays might have durations comparable 
to those observed in the rest of the sample.

Burst 58 has a much dimmer second peak with respect to the other bursts,
only about 15 per cent above the net burst count rate,
which is however still significantly above the persistent emission.
It features the longest dip, at $\Delta t\sub{dip} =72$~s,
going down to just about 4 per cent above the persistent emission,
but still significantly above the noise fluctuations.

The recurrence time for bursts 118 and 119 is 1.8~ks,
similar to what is observed for bursts 35--37.
Their dips are about 30~s long, and briefly drop to count rates within 1 per cent
of those observed right before the bursts.
Like in burst 37, the QPO remains visible during the dip.
Their second peaks are about 30 per cent as large as the first ones.

		\subsubsection{Spectral properties of the pre-burst emission}\label{sec:perssp}

\begin{table*}
\begin{minipage}{2.\columnwidth}
 \caption{Properties of the double-peaked bursts}\label{tab:spectra}
 \begin{tabular}{@{}lccccccc}
   \hline\hline
Burst	&	Obs ID	&	$F\sub{peak,1}^{(a)}$	&$F\sub{peak,2}^{(b)}$	&	$E\sub{b}^{(c)}$&$E\sub{sf}/E\sub{b}^{(d)}$&	$\Delta t\sub{dip}^{(e)}$& $D^{(f)}$	\\
		&			&	[\ergscm{-9}]	&	[\ergscm{-9}]		&	[\ergcm{-8}] &		& [s]				& \\
   \hline
35	&	30079-01-03-00R	&	6.0$\pm$0.5	&	2.3$\pm$0.6	&	23.4$\pm$1.1	&	$(35\pm6)$\%	&	46	&	44\%	\\%&	173
36	&	30079-01-03-00R	&	5.7$\pm$0.4	&	1.8$\pm$0.4	&	20.7$\pm$1.1	&	$(34\pm7)$\%	&	46	&	56\%	\\%&	165
37	&	30079-01-03-01R	&	5.9$\pm$0.5	&	2.7$\pm$0.6	&	28.8$\pm$1.7	&	$(31\pm9)$\%	&	25	&	11\%	\\%&	193
58	&	40433-01-02-00R	&	9.5$\pm$0.5	&	1.1$\pm$0.3	&	29.9$\pm$1.5	&	$(73\pm7)$\%	&	72	&	26\%	\\%&	134
118	&	95410-01-01-00	&	5.2$\pm$0.8	&	1.2$\pm$0.3	&	17.1$\pm$1.9	&	$(28\pm6)$\%	&	33	&	17\%	\\
119	&	95410-01-01-00	&	5.4$\pm$0.8	&	1.2$\pm$0.3	&	19.1$\pm$2.1	&	$(26\pm4)$\%	&	30	&	16\%	\\
   \hline

 \end{tabular}

 \medskip
All fluxes and fluences are bolometric. $^{(a)}$ First-peak flux. $^{(b)}$ Second-peak flux. $^{(c)}$ Burst fluence. $^{(d)}$ Fractional fluence before the dip.
$^{(e)}$ Dip duration. $^{(f)}$ Dip depth, the fraction of photons lost to the dip. See Sec.\ref{sec:obsana} for explanation.
\end{minipage}
\end{table*}

Burst 35 and 36 were observed during a direct pointing,
while burst 37 took place in an offset configuration (see Sec.\ref{sec:obsana}).
The pre-burst spectrum of the RB was derived from the available 182~s-long
segment of the lightcurve prior to burst 37.
We fitted this with a model that is a good description
of all RB non-burst data (B13)
and consists of a disc black body \citep{1984PASJ...36..741M},
a power law and a Gaussian line at 6.4~keV.
Low-energy absorption by the interstellar medium was taken into
account using the model of \citet{1983ApJ...270..119M},
with an equivalent hydrogen column density of $N_{\rm H}=1.6\times10^{22}$~cm$^{-2}$
(\citealt*{1995AJ....109.1154F}; \citealt{2000A&A...363..188M}; \citealt{2004A&A...426..979F}).
The resulting blackbody temperature is $(2.48\pm0.06)$~keV,
the apparent disc radius $(1.77\pm0.06)\textrm{~km}\times (D/\textrm{10~kpc})/\sqrt{\cos i}$
\citep[with $D$ the source distance and $i$ the disc inclination;
for the relation between the apparent and realistic inner disc radius, see][]{1998PASJ...50..667K}
and the power-law photon index is $2.35\pm0.23$.
The goodness of the fit is $\chi^2\sub{red} = 1.25$ for 46 degrees of freedom (dof).
We call this the ``RB spectrum''.

For the persistent emission prior to bursts 35 and 36,
whose observations are contaminated by the nearby source 4U 1728-34,
which dominates above 20~keV,
simply renormalizing the RB spectrum according to the collimator responses
(1.0 and 0.42 for the on- and offset pointings respectively)
while leaving the other parameters frozen
(under the assumption that the spectrum does not change significantly
over the short time separating the observations) yields unacceptable fits
[for burst 35, $\chi^2\sub{red}\textrm{(dof)} = 53.2$(51)].
We therefore also added an absorbed Comptonization model
($N\sub{H} = 2.4 \times 10^{22} \textrm{~cm}^{-2}$; \citealt{2011A&A...530A..99E}),
which we refer to as the ``4U 1728-34 spectrum''.
This fit gave $\chi^2\sub{red}\textrm{(dof)} = 1.10$(46) and 1.04(46)
for the persistent-emission spectrum before bursts 35 and 36 respectively.

Analogously, a persistent emission spectrum was extracted
from an offset pointing 1200~s after burst 58,
corrected for the collimator response,
and added to a component describing 4U 1728-34,
for an overall good fit with $\chi^2\sub{red}\textrm{(dof)} = 0.85$(42).
Finally, no offset observations were available near bursts 118 and 119.
Therefore, we simply fit a model similar to the one described above
(a ``RB'' plus a ``4U 1728-34'' component) to each of their pre-burst spectra,
which gave $\chi^2\sub{red}\textrm{(dof)} = 0.63$(49) and 0.88(49) respectively.

			\subsubsection{Spectral properties of the bursts}\label{sec:spe}

\begin{figure}
  	\includegraphics[width=0.95\columnwidth]{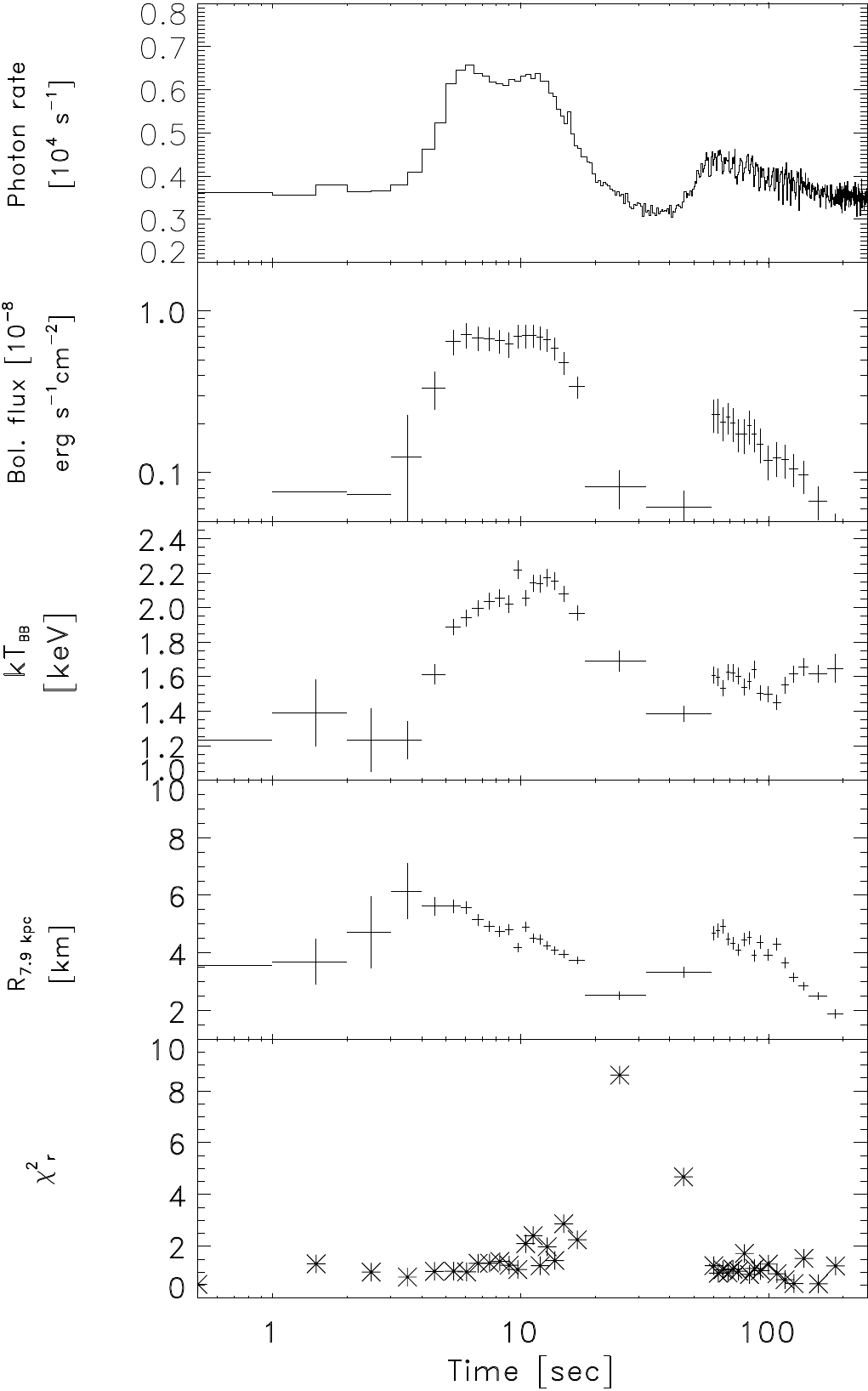}
	\caption{\small{Time-resolved spectroscopy of burst number 35.
The top panel shows the measured full-bandpass photon count rates for the PCA
Xenon layers.
Following downward, the bolometric flux, the blackbody temperature and radius
(assuming a distance of 7.9~kpc) and the $\chi^2\sub{red}$ of the fit are plotted.
It is clear that no PRE takes place,
and that during the dip the pre-burst spectrum
yields an unacceptable fit (see Sec.~\ref{sec:res}).}}
	\label{fig:35sp}
\end{figure}

Spectra during the bursts were modelled by the combination of black
body radiation (leaving free temperature and emission area, to account
for the varying burst emission, and multiplied by a model for interstellar absorption
with $N\sub{H} = 1.6 \times 10^{22} \textrm{~cm}^{-2}$)
and the model as found for the non-burst spectrum discussed above,
keeping the parameters of the latter fixed.
The spectral properties of the bursts are listed in Table~\ref{tab:spectra}.

Bursts 35 to 37 reach remarkably similar bolometric peak fluxes
$F\sub{peak}$ (in units of \ergscm{-9}),
from 5.7$\pm$0.4 to 6.0$\pm$0.5 in the first peak,
while ranging from 1.8$\pm$0.4 to 2.7$\pm$0.6 in the second one.
Burst 58 has instead a 50 per cent brighter first peak,
at $F\sub{peak} =$ 9.5$\pm$0.5,
and a dimmer second peak, at 1.1$\pm$0.3.
Finally, bursts 118 and 119 have bright first peaks,
comparable to those of bursts 35--37,
and dim second peaks like that of burst 58.
Such fluxes are all significantly sub-Eddington:
for the quoted distance, 
the Eddington flux
is $F\sub{Edd} =$ 2.1$\times$\ergscm{-8}
for a ``canonical'' NS with mass $M = 1.4 M_{\odot}$
accreting solar-composition material (hydrogen fraction $X=0.7$).

The total burst fluence $E\sub{b}$
spans from (17.1$\pm$1.9)$\times$\ergcm{-8} to (29.9$\pm$1.5)$\times$\ergcm{-8}.
Despite its brighter first peak, burst 58 is not more energetic than burst 37.
This is again because there are differences
in how the energy is distributed before and after the second peak:
in bursts 35--37, 118 and 119,
27 to 35 per cent of the fluence is emitted before the dip,
and 65 to 73 per cent after it.
In burst 58 instead,
73 per cent of the fluence is emitted up to the dip minimum
and only 27 per cent after it.

The results of the time-resolved spectroscopic analysis for burst 35
are plotted in Fig.~\ref{fig:35sp}.
Two main conclusions can be drawn.
Firstly, a decreasing temperature accompanied by an increasing radius
while the flux drops out of the instrumental bandpass,
the hallmarks of photospheric expansion
(see e.g. \citealt{2008ApJS..179..360G}),
are not visible in the spectral evolution.
This is also true during the other dips.
Therefore, we conclude that they are not due to PRE.

Secondly, the goodness of fits worsens during the decay into the dips of bursts 35 and 36,
until $\chi^2\sub{red}$(dof) becomes as large as 8.5(27) and 4.2(26) respectively
when the count rate is roughly at the pre-burst level.
We therefore examined whether the spectrum in these dips
is different from that in the persistent emission prior to the burst.

We attempted to fit the dip spectrum of burst 35
with three different models.
The spectrum was extracted between 32 and 40~s after the burst start,
when the flux is stable at about 5 per cent below the count rate
of the persistent emission prior to the burst (see Fig.~\ref{fig:humps}).
The model that fitted the pre-burst emission (see Sec.\ref{sec:perssp})
was unacceptable for the burst dip, with 
$\chi^2\sub{red}\textrm{(dof)} = 6.68$(29).
Letting the two additive components
that make up the persistent emission
(the RB and 4U 1728-34 spectra)
free to vary in normalization also yielded an unacceptable fit,
with $\chi^2\sub{red}\textrm{(dof)} = 4.18$(27).
This means that the fit cannot be improved by simply
allowing the normalization of the subtracted pre-burst spectrum vary,
as successfully done by \citet*{2013ApJ...772...94W} for PRE bursts.
The energy dependence of the ratio of the model to the data
rather suggests that the discrepancy results from
an intrinsically different spectral shape.

We therefore fitted the dip by
keeping the 4U 1728-34 spectrum frozen and
leaving free to vary different components of the RB spectrum.
The only solution with an acceptable fit
required a column density $N\sub{H} =$ (6.86$\pm$0.96)$\times 10^{22} \textrm{~cm}^{-2}$,
more than four times larger than the
quoted value, which we normally employ for the pre-burst emission.
The required blackbody temperature and radius were
(2.03$\pm$0.06)~keV and  (2.55$\pm$1.04)~km.
This choice gave the best-fit model,
with $\chi^2\sub{red}\textrm{(dof)} = 1.28$(26).
Conversely, it was not applicable to the pre-burst emission,
where it yielded $\chi^2\sub{red}\textrm{(dof)} = 96.4$(52).

As a final test, we extracted the spectrum of the persistent emission
1000~s after the burst start.
In this stretch of data, the count rate is on average 2.5 per cent
below the pre-burst emission.
We tried to fit the spectrum of this post-burst emission
with both the model for the pre-burst emission and for the dip,
letting the ``RB spectrum'' free to vary in normalization
while freezing all the other components.
The pre-burst model gave $\chi^2\sub{red}\textrm{(dof)} = 1.32$(51)
with a normalization 4.2 per cent lower,
while the model that best fit the dip resulted
in a $\chi^2\sub{red}\textrm{(dof)} = 79.07$(51),
with a 2.5 per cent higher normalization.

Therefore, the dip has a truly different spectral shape
than both the pre- and post-burst emission,
even though they are roughly at the same count rate.

	\subsection{Timing analysis}\label{sec:QPO}

\begin{table*}
\begin{minipage}{2.\columnwidth}
 \caption{Log of the \textit{RXTE} timing measurements of the RB power spectral fit parameters}\label{tab:QPO}
 \begin{tabular}{@{}lcccccccc}
   \hline\hline
Obs ID	&	date	&	start time &	$t\sub{obs}$ &	rms	&	Q	&	$\nu_0$	&	significance	&	count rate \\ %&$chi^2\sub{red}$(dof)
	&		&	(UTC)	   &	[ks]	&	[\%]	&		&	[Hz]	&	$\sigma$	&	[c/s/PCU] \\
   \hline

%30078-01-08 &	1998 Aug 19	&	03:16:59	&	4.2	&	1.03 $^{+ 0.66 }_{- 0.62 }$ &	4.95$\pm$3.78	&	0.22$\pm$0.01	&	2.55	&	354 \\	%&1.03(96)	-> 0.39
%30079-01-02 &	1998 Aug 19	&	04:56:44	&	3.8	&	1.49 $^{+ 0.90 }_{- 0.98 }$	&	2.35$\pm$1.40	&	0.28$\pm$0.03	&	2.58	&	333 \\	%&0.95(107) -> 0.61
30079-01-03$^{a}$ &	1998 Aug 22	&	08:10:36	&	8.7	&	3.93 $^{+ 0.95 }_{- 0.97 }$	&	2.29$\pm$0.21	&	0.25$\pm$0.004	&	16.64&	569 \\	%&1.15(106) -> 0.137
40058-02-03 &	1999 Mar 19	&	18:12:43	&	17.2	&	1.91$^{+ 0.84 }_{- 0.82 }$&	1.42$\pm$0.43	&	0.20$\pm$0.016	&	5.33	&	320 \\	%&0.92(108)	-> 0.71
40433-01-02$^{b}$ &	1999 Oct 08	&	12:06:23	&	3.0	&	2.28$^{+ 0.67 }_{- 0.63 }$	&	0.43$\pm$0.13	&	0.20$\pm$0.017	&	12.13	&	555 \\	%&0.92(118)	-> 0.89
50420-01-09 &	2001 Feb 16	&	14:12:45	&	11.7	&	5.03$^{+ 1.47 }_{- 1.48 }$ 	&	0.79$\pm$0.10	&	0.26$\pm$0.011	&	11.68	&	266 \\	%&1.19(108)	-> 0.087
95410-01-01$^{c}$ &	2010 Jan 15	&	20:41:33	&	2.8	&	2.61$^{+ 1.06 }_{- 1.11 }$	&	2.01$\pm$0.66	&	0.18$\pm$0.006	&	5.74	&405 \\	%&0.82(109)	-> ok

   \hline

 \end{tabular}

 \medskip
Errors on the parameters were determined using $\Delta \chi^2= 1$.
$^{a}$Includes bursts 35 to 37. $^{b}$Includes burst 58. $^{c}$Includes bursts 118 and 119.
\end{minipage}
\end{table*}

A timing analysis of a dataset
which included the double-peaked bursts 35 through 37
was already performed by \citet{2001MNRAS.321..776F}.
However, their main focus was the search of high-frequency burst oscillations,
and they calculated Fourier transforms of 1~s long data stretches,
thereby ignoring the entire sub-Hz portion of the power spectrum.

For the timing analysis we used all Good Xenon and Event data,
with a time resolution of $2^{-20}$ and $2^{-13}$~s, respectively.
The Good Xenon data were rebinned to the same resolution as the Event data.
No background subtraction or dead-time correction was applied to the data
before calculating the power spectra,
but type I and II bursts were removed.
Following a well-established procedure (see e.g. \citealt*{2004NuPhS.132..381K})
we determined the Poisson noise using the formula by \citet{1995ApJ...449..930Z}
and then shifted it to the featureless level
between 3072 and 4096 Hz,
a region in the power spectra dominated by photon counting noise.
This mean value was then subtracted from the power spectra.
The background was calculated for each observation (Obs ID)
with the \textsc{ftool pcabackest} over the entire energy range.
We used 128~s-long segments to calculate power spectra with no energy selection.
The frequency boundaries for the power spectra are therefore 1/128 Hz and 4096 Hz.
The power spectra were normalized to the relative rms \citep{1995lns..conf..301V}
which yields the power density in units of (rms/mean)$^2$~Hz$^{-1}$.

\begin{figure}
  	\includegraphics[width=0.95\columnwidth]{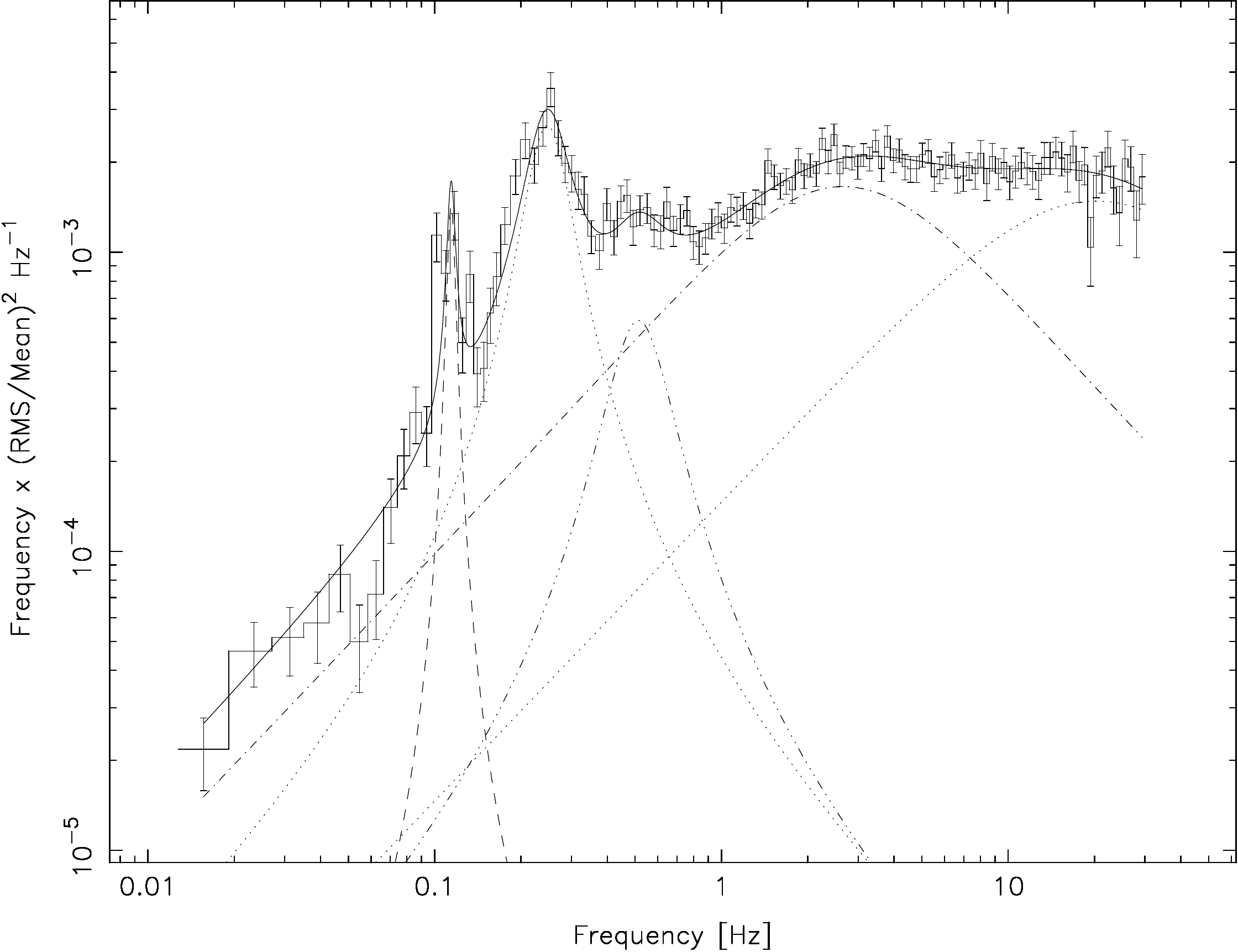}
	\caption{\small{Logarithmically-rebinned, power spectrum distribution
 of all the non-burst data in Obs ID 30079-01-03,
which features bursts 35 to 37.
Notice the QPOs at 0.1 and 0.25~Hz.
This is the only detection of the former feature and
the most significant detection of the latter.
The first harmonic of the 0.25~Hz QPO is also visible at 0.5~Hz.}}
	\label{fig:powsp}
\end{figure}

\begin{figure}
  	\includegraphics[width=0.95\columnwidth]{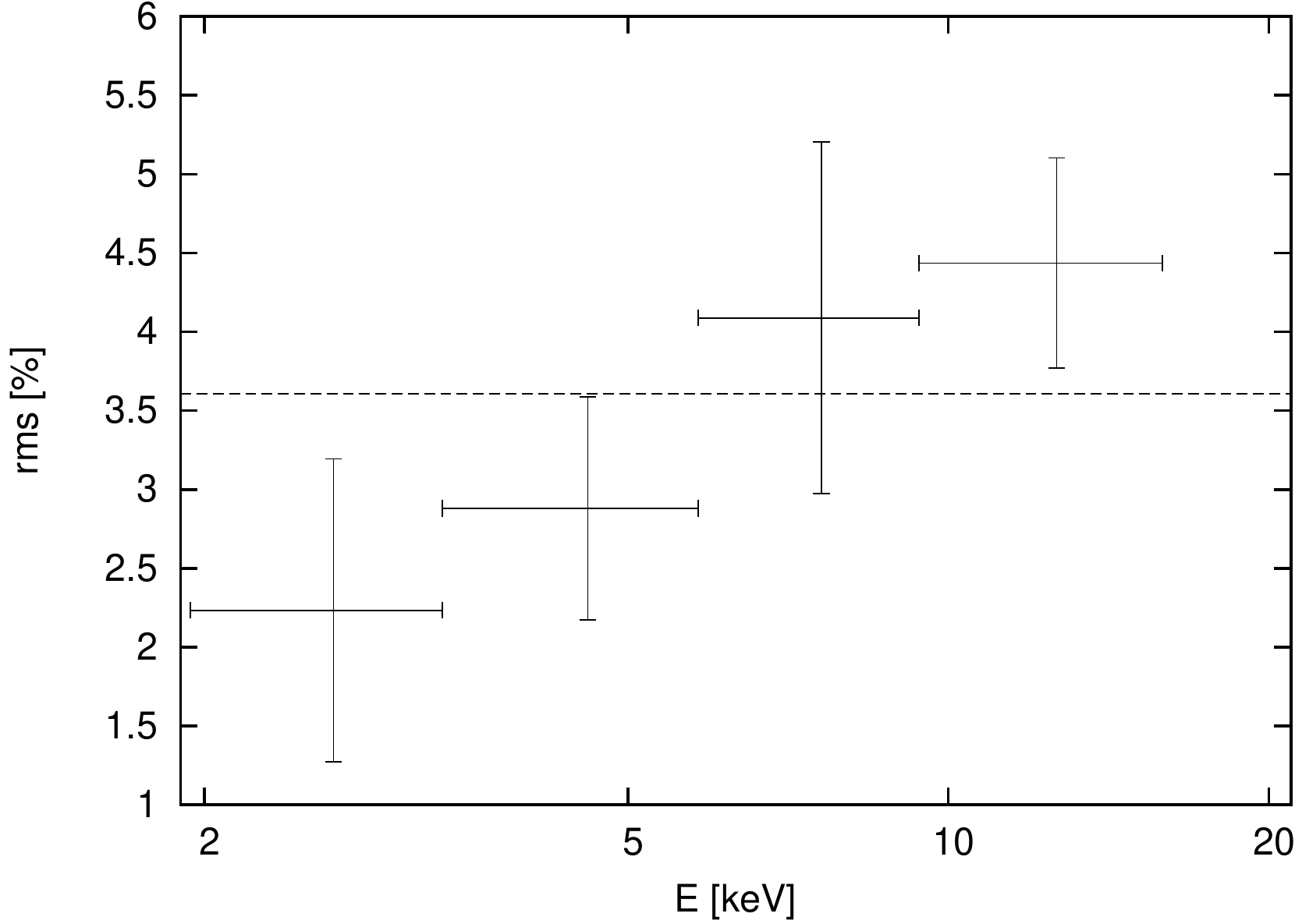}
	\caption{\small{The energy dependence of the $\sim 0.25$~Hz QPO,
over the same data stretch as covered by Fig.~\ref{fig:powsp}.
The rms increases with energy.
The plot show rms amplitudes up to 17~keV,
where counts yield acceptable fits to the feature.
Error bars correspond to $1 \sigma$ levels,
the dashed line shows the average rms over all energies.}}
	\label{fig:rms_vs_E}
\end{figure}

We scanned all the PCA power spectra.
Each power spectrum was inspected by eye to look for the presence of the LF QPO.
When it is present,
we fit the logarithmically-sampled low frequency ($\nu < 30$~Hz)
portion of the spectrum with a sum of Lorentzian components,
following \citet*{2002ApJ...572..392B}.
Each Lorentzian peaks at a frequency 

\begin{equation}
 \nu\sub{max} = \nu_0 \sqrt{1 + \frac{1}{4Q^2}},
\end{equation}
with centroid frequency $\nu_0$
and coherence factor $Q$ corresponding to the ratio of $\nu_0$ to the FWHM.
An example of these fits can be seen in Fig.~\ref{fig:powsp}.
We then integrated the fractional rms of each Lorentzian over the whole frequency range,
and over smaller energy bands,
to quantify the energy dependence of the rms (see Fig.~\ref{fig:rms_vs_E}).
Table~\ref{tab:QPO} lists all observations
in which the 0.25~Hz QPO was found, and the QPO properties.

We detect the QPO in five distinct Obs IDs,
across five different outbursts.
They all feature either double-peaked type I bursts
or the kind of type II bursts (long, bright, flat-topped)
which are the first to appear and are only present in a short phase of an outburst decay
(see Sec.~\ref{sec:outb}).
The count rates vary by a factor of two across these observations.
The fits yielded acceptable $\chi^2\sub{red}$(dof),
varying between 0.92(118) and 1.19(108),
across the different observations.
The significance of the detection, defined as the ratio of
the integrated power over its negative error
varies between 5.33 and 16.64~$\sigma$.

The power spectrum of the observation which includes
bursts 35 through 37 is shown in Fig.~\ref{fig:powsp}.
This is the most significant detection of the 0.25~Hz QPO.
Besides the main feature an accompanying QPO is visible at 0.1~Hz.
Also present are QPOs that peak at 0.5 and 3~Hz,
which are known to appear in the run-up to a large type II burst \citep{1988ApJ...324..379S}.
The former is probably in a harmonic relation to the 0.25~Hz QPO.

Our newly discovered QPO, relatively well fit by a Lorentzian,
covers a relatively narrow frequency range in the different observations,
between 0.18 and 0.26~Hz.
Its rms amplitude varies between (1.91$\pm$0.83) to (5.03$\pm$1.47) per cent.
At its peak, it is also clearly observed in the time domain (see Fig.~\ref{fig:humps}).
As can be seen in Fig.~\ref{fig:rms_vs_E},
the amplitude seems to grow monotonically over the 2--17~keV
(channels 5-64 of the PCA) energy range.
The error bars are however quite large,
and a constant value could also fit the rms-$E$ relation.
We performed an F-test to compare the fits
for a constant value and a linear dependence on energy.
The linear function yields a better fit,
and the probability of the improvement being due
to random chance is $p=0.04$.
We therefore consider the growing trend real.
The coherence factor $Q$ is rather low, between 0.43 and 2.29.
However, we must stress that the feature is sometimes observed
to increase in frequency during the intensity decrease
that precedes the longest type II burst.
This degrades its coherence in our fit,
which is averaged over an entire observation.

The weaker 0.1~Hz QPO feature 
was only detected on the same date as the double-peaked bursts 35--37,
while it is not present in the other power spectra,
and it is poorly fit by a Lorentzian.
Therefore, we cannot establish whether it might be in a harmonic relation
with the 0.25~Hz oscillation, nor can we exactly quantify its amplitude or coherence.
It is however quite a narrow feature, and it seems weaker than the 0.25~Hz QPO.

We did not fit the power spectra of the remainder of the observations,
but a visual inspection of the power spectra indicated that the QPO disappears
as the rare, 300~s long, flat-topped type II bursts
(which are the first to appear in an outburst, \citealt{1999MNRAS.307..179G})
evolve into shorter, more frequent ones at lower count rates.

	\subsection{The colour-colour diagram}

There are $\simeq$335~ks of non-contaminated exposures
with 4U 1728-34 outside the field of view and the active RB inside.
We employed 16~s time resolution \textsc{standard 2} data
for these observations to calculate X-ray colours.

Following \citet{2010ApJ...719..201H},
soft and hard colours are defined as the ratio of counts
in the $\simeq$4.1-7.3~keV to that in the $\simeq$2.5-4.1~keV band,
and in the $\simeq$9.8-18.1~keV and $\simeq$7.3-9.8~keV bands, respectively.
Type I and II bursts were removed, background was subtracted,
and dead-time corrections were made.
The energy-channel conversion was determined from the pca\_e2c\_e05v01 table
provided by the \textit{RXTE} team.
In order to correct for gain changes
as well as for differences in effective area between the PCUs,
we normalized our colours to the corresponding Crab source colour values
\citep*{2003ApJ...596.1155V} that are
closest in time and in the same \textit{RXTE} gain epoch,
to make sure the same high-voltage settings of the PCUs apply.
Continuous stretches of data points were combined
until the relative error of each colour was below 5 per cent.
Finally, we flagged the colours closest in time to the double-peaked bursts 37 and 58.
The former measurement is also very close to bursts 35 and 36.
Bursts 118 and 119 have no close-by non-contaminated exposure.

\begin{figure}
	\includegraphics[width=.95\columnwidth]{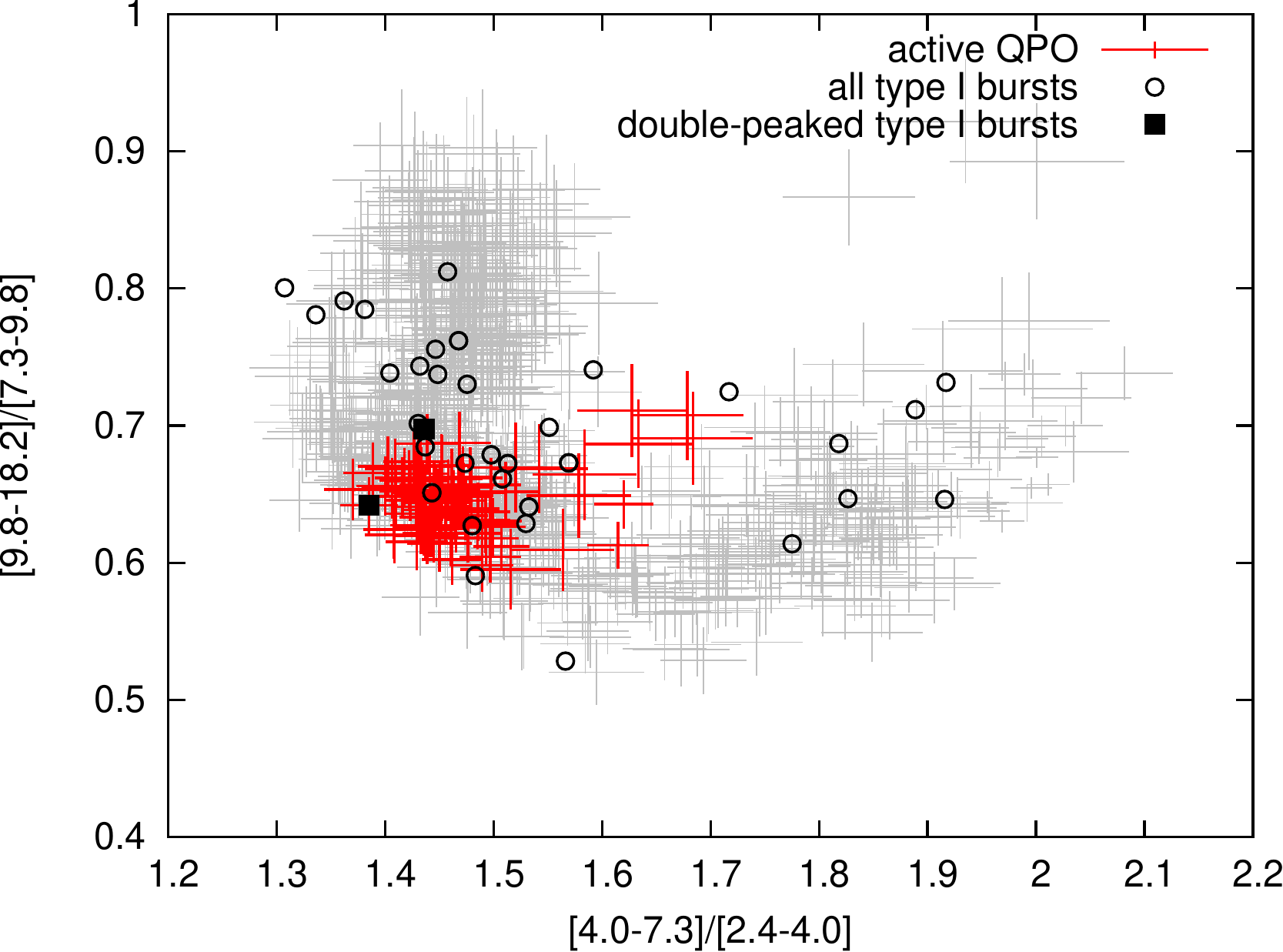}
	\caption{\small{Colour-colour diagram (CCD)
of the RB.
Circles indicate colours of the persistent emission as close to a type I burst as possible,
filled squares indicate the same but for four double-peaked bursts.
The red data points refer to the data where the QPO is observed.
While ordinary type-I bursts take place in all source states,
the double-peaked bursts and the 0.25~Hz QPO
are only visible at the vertex between the island and banana branches.
Offset observations were chosen, for a total of 335~ks,
so that 4U 1728-34 is not contaminating
the field of view of the PCA, and
type I and II bursts were removed from the data.
Each data point represents 16~s of data.
Where needed, the data were binned in time, to keep errors on each colour below 5 per cent.}}
	\label{fig:CD}
\end{figure}

Fig.~\ref{fig:CD} shows the resulting colour-colour diagram (CCD).
Two clearly distinct branches are present
which indicate that the RB is an atoll source,
with the so-called banana and island states \citep{1989A&A...225...79H}.
Type I bursts take place in a wide range of mass accretion rates,
as already shown by B13.
Interestingly, the double-peaked bursts
and the accompanying 0.25~Hz QPO concentrate at the vertex where the RB
transits from the lower banana to the island state.

\section{Discussion}\label{sec:dis}

\subsection{Observational summary}

In a comprehensive analysis of all \textit{RXTE} data on the RB
we found six type I X-ray bursts during three outbursts
that feature a dip that is not due to photospheric radius expansion,
resulting in the appearance of two peaks in the light curves.
The dips are relatively long with durations between 25 and 72 s
and ingresses and egresses that have a typical time scale of 10 s.
In two bursts, the dips drop for about 10~s
below the pre-burst flux level, by up to 5\%.
The dip is therefore also in the persistent emission,
although never outside a burst.
When that happens, the spectral shape is different from the pre-burst spectrum
with an increase of the absorption column density by a factor of 4.
After the burst, the persistent emission recovers
its pre-burst spectral shape.
Such a peculiar and repetitive morphology of type I burst light curves has,
to the best of our knowledge, never been observed before (see Sec.~\ref{sec:multipeak}).

With respect to the other type I bursts from the RB,
the duration of the double-peaked bursts seems longer 
and the decays after the second peak brighter,
but that may be due to the addition of the QPO
on the tail and not on the first peak.
The fluence of the bursts ranges between 17.12 and
28.80$\times$\ergcm{-8}. This compares with 17.6
to 51.5$\times$\ergcm{-8} for the other 121 bursts
(B13).
If the dips hide part of the actual fluence,
the total fluence would be between 10 and 50\% larger,
varying between about 20 and 37$\times$\ergcm{-8}.
Instead, the pre-dip fluence alone is between 4.80 and
9.01$\times$\ergcm{-8} which is significantly below the
lowest value for all other bursts.
These ranges are a strong indication that
the second peaks are thermonuclear in nature and not,
for instance, completely due to accretion.
The other burst properties, like peak flux,
rise and recurrence times
are quite unremarkable compared to the rest of the sample.
However, it must be noted that the recurrence time of the bursts
spans a rather limited range, between 1.7 and 2.2~ks,
and that they appear at similar persistent count rates.
B13 showed that the RB produces bursts over a very large range
of persistent fluxes, and that the recurrence time scales accordingly.
It seems therefore likely that the double-peaked bursts are peculiar
to a limited range of mass accretion rates.

\textit{RXTE} data of the RB suffer from contamination of another bright LMXB
in the same field of view: 4U 1728-34. This does not affect the
analysis of X-ray bursts since those of the RB are easily discerned
from those of 4U 1728-34, but the analysis of the non-burst radiation
is seriously affected, which precluded thus far the identification of the
RB as an atoll or Z-source \citep{1995MNRAS.277..523R}.
However, a portion of the observations were pointed such that 4U 1728-34 is
outside the field of view, while the RB remains inside, though at a
less favorable off-axis angle.
Thanks to the large size of the dataset collected by \textit{RXTE} during its mission,
we generated a CCD from these data and are able, for the first time,
to identify the RB as an atoll source
with unambiguous `island' and `banana' branches \citep{1989A&A...225...79H}.

The double-peaked bursts seem to occur at a phase in the decay of the
transient outburst when type II bursts commence, as signalled by the
presence of very energetic, flat-topped and long type II bursts \citep{1999MNRAS.307..179G}.
Furthermore, the double-peaked bursts are coincident
with a special location in the CCD - the vertex where the source
transits between the banana and the island state,
indicating a change in the accretion-flow geometry \citep{2006csxs.book...39V}.

The non-burst light curve also shows a peculiar feature,
around the double-peaked bursts: a QPO with a typical frequency of about 0.25 Hz.
This QPO is present before and after the bursts,
and during the second peak of the bursts.
The amplitude peaks at 5.03$\pm$1.47\% fractional rms.
Notably, it is absent during the first burst peak and the deep dips.
It remains visible during shallow dips with a smaller amplitude.
When compared to other type I bursts from the RB
and considering the depth of the deepest dips,
this strongly suggests that the QPO represents an additional radiation component
to the burst and the persistent emission.

We searched all \textit{RXTE} data on the RB
and found a total of five statistically significant detections
(see Sec.~\ref{sec:QPO}).
The feature is present in all the consecutive
power spectra that make up each observation
and it also always lies in the same portion of the CCD
across five different outbursts,
at the transition between the island and banana states.
Furthermore, all the observations featuring the 0.25~Hz QPO
coincide with either double-peaked bursts or
the kind of type II bursts (long, bright, flat-topped)
which only appear in a short-lived phase of an outburst decay
\citep{1999MNRAS.307..179G}.

The observations of mid-burst dips during type-I X-ray bursts
and the simultaneous behavior of the 0.25 Hz QPO
at a state transition are unique.
How do they come about and how are they related to each other?
Why do they take place during a state transition?
Are they related to the unique type II bursting behavior of the RB and, if so,
can they provide us further constraints on the origin of that behavior?

\begin{table*}
\begin{minipage}{2.\columnwidth}
 \caption{Mechanisms for the dips and QPO}\label{tab:mechanisms}
 \begin{tabular}{@{}lcccccccc}
   \hline \hline
Model	&	\multicolumn{3}{c}{Bursts}		&	\multicolumn{3}{c}{QPO}		&	\multicolumn{2}{c}{Accretion state}\\
 	&	Double peaks	$^{(a)}$&	$\Delta t\sub{dip}^{(b)}$	&	Spectrum$^{(c)}$&	$\nu_0$	&	rms	&	rms/Energy$^{(d)}$	&	Soft/Hard$^{(e)}$	& Type II burst$^{(f)}$	\\
   \hline
 Unburnt layer			&	\textbf{Y}	&	P	&	N	&	...	&	...	&	...	&	...	&	...\\
 Reaction chain stalling	&	\textbf{Y}	&	N	&	N	&	...	&	...	&	...	&	...	&	...\\
 Equatorial stalling		&	\textbf{Y}	&	P	&	N	&	...	&	...	&	...	&	P	&	...\\
 Magnetic confinement		&	\textbf{Y}	&	P	&	N	&	...	&	...	&	...	&	N	&	...\\
 Burst-induced corona		&	N		&	N	&	\textbf{Y}	&	...	&	...	&	...	&	...	&	...\\
 Marginally-stable burning	&	...		&	...	&	...	&	N	&	\textbf{Y}	&	N	&	\textbf{Y}	&	...\\
 High inclination (dipper)	&	\textbf{Y}	&	\textbf{Y}	&	\textbf{Y}	&	\textbf{Y}	&	\textbf{Y}	&	N	&	N	&	...\\
 Disc-magnetosphere instability	&	...	&	...	&	...	&	\textbf{Y}	&	N	&	\textbf{Y}	&	N	&	\textbf{Y}\\
 Low inclination, two channels	&	\textbf{Y}	&	P	&	\textbf{Y}	&	P	&	...	&	...	&	\textbf{Y}	&	P\\
   \hline

 \end{tabular}

 \medskip
The table compares the properties of the bursts, the QPO and the source state
against the predictions made by the various models discussed in Sec.\ref{sec:dis}.
Y/N
(yes/no) indicates that the model can/cannot explain the property in a given column.
The symbol P (possible) indicates that the model might be able to explain the property if certain conditions are
met.
An empty field indicates that the model makes no specific prediction for that property.
$^{(a)}$Non-PRE, double-peaked bursts; 
$^{(b)}$dip duration; 
$^{(c)}$enhanced absorption in the dip spectrum and dip going below the persistent flux level;
$^{(d)}$QPO fractional amplitude growing with energy;
$^{(e)}$appearance of the dip and the QPO only at the soft/hard transition;
$^{(f)}$appearance of the dip and QPO at the same time as the first type II bursts.
\end{minipage}
\end{table*}

\subsection{Nature of the dips}\label{sec:multipeak}

Other LMXBs exhibit double-peaked bursts 
that are not due to PRE.
For example 4U 1636-53 shows bursts
with two \citep{1985ApJ...299..487S,2006ApJ...636L.121B,2006ApJ...641L..53B}
and even three peaks \citep{1986MNRAS.221..617V,2009MNRAS.398..368Z}.
However, the time scales
for this source are different than for the RB (the peaks are separated by
at most 5 s) and the flux never drops below the pre-burst
emission. The similarities are that the peak flux is also a factor of
2 to 3 below the Eddington limit, the inferred mass accretion rate is
relatively large and the bursts concentrate around the vertex of the
island/banana branch of the CCD.
Non-PRE double-peaked type I bursts have also been observed from
4U 1608-52 \citep{1989A&A...208..146P},
GX 17+2 \citep{2002A&A...382..947K}
and 4U 1709-267 \citep{2004MNRAS.354..666J},
all with durations shorter than 20~s.

There is great diversity in the observations of other sources,
precluding a single explanation for all irregularities.
The double-peaked bursts in the RB are no exception:
they too are unique, particularly because of relatively long time
scales for the dip ingress and egress and the simultaneous QPO behavior.
We will now discuss several explanations
for the double-peaks phenomenon in the other sources.

	\subsubsection{Two-stage energy release on the surface?}

One possibility is that the multiple peaks are
the result of a variation in either the rate
at which the energy is transported to the surface of the NS,
or in the rate at which energy is released in the flash.
A shear instability could mix layers
which have undergone a thermonuclear flash
with unburnt material lying above them \citep{1988A&A...199L...9F},
or there could be a waiting point in the rp-processes \citep*{2004ApJ...608L..61F}.
However, it is difficult to keep unburnt fuel above burnt material
without the two mixing \citep*{2002ApJ...566.1018S},
or to have the burning reaction chain
stalling as long as the dips are sometimes observed to last \citep{2006ApJ...636L.121B}.

Other explanations discussed in literature involve
stalling of the burning front.
For a weak burst with high-latitude ignition,
the spreading of the flame could stall at the equator
if that is where most of the disc material is accreted
\citep{2006ApJ...636L.121B}.
Otherwise, a temporary boosting of the magnetic field $B$,
due to the motion of the fluid upon ignition,
could temporarily confine the burning region
\citep{2011ApJ...740L...8C}.
Obviously $B$
is not expected to be different
for the double-peaked bursts and the rest of the RB bursts,
nor is the fuel composition, given their similar timescales and energetics.
The ignition location and the mass-accretion rate
are therefore the remaining possible triggers for a confining mechanism
that could give rise to the double peaks.
Perhaps high-latitude ignition could be rare enough
to only happen in 6 bursts out of 125,
although these are not among the weakest in the sample
as observed for double-peaked bursts in 4U 1636-53 \citep{2006ApJ...636L.121B}.
On the other hand, the possibility of magnetic confinement
is at odds with the fact that the RB double-peaked bursts
occur in a limited range of intermediate mass-accretion rates,
with single-peaked bursts being observed at both higher and lower $\dot{M}$.

	\subsubsection{Transient obscuration from the disc?}

Generally speaking, all the scenarios discussed above
revolve around phenomena confined to the surface.
None of them can therefore explain why, in bursts 35 and 36,
the flux drops below pre-burst levels
while the absorption column density increases.
The accretion flow, responsible for the non-burst emission, must clearly also be affected.

For example, the multi-peaked structure could result
from scattering of the X-rays by material evaporated from the accretion disc
during the flash, forming a burst-induced accretion disc corona.
Models seem however to require Eddington-limited bursts \citep{1992ApJ...398L..53M},
and indeed observations of radiative or dynamical interactions with the accretion disc
have so far been limited to very energetic PRE bursts
\citep{2011A&A...525A.111I,2013A&A...553A..83I,2013ApJ...767L..37D}.
Instead, the double-peaked bursts in the RB
are sub-Eddington by at least a factor of 2,
implying that the burst flux delivers insufficient radiation pressure to
influence the accretion flow.
Even resorting to beaming,
to raise the flux in a direction
outside the line of sight to a sufficient level,
the problem remains
that the time scales are dynamical in nature and
much shorter than those observed \citep[e.g.][]{1992ApJ...398L..53M}.

The energetic properties of these bursts
are unremarkable with respect to the rest of the type I sample,
which makes explaining their exceptionality difficult.
None the less, the dips seem to be initiated by the bursts,
responding to the burst with a delay of about 10~s.
As a way around this, the origin of the dips
might lie in a short-lived,
peculiar geometry of the inner accretion flow
rather than in the bursts themselves.

For a scenario that is independent of burst properties
and features longer dips, one can look at dipping sources.
These show periodic dips in their light curves
that are thought to be caused by periodic obscuration
of the central source by a structure formed in an interplay
between the gas stream from the companion and the accretion disc \citep{1982ApJ...253L..61W}.
Therefore, they have generally been inferred to 
be at inclination angles larger than $60\degr$ \citep*{1987A&A...178..137F},
unless the splash traverses over the compact object to the
other side of the accretion disc
\citep[e.g.,][]{1998ApJ...493..898A}.

The depth, duration and spectral properties of the dips vary from
source
to source and from cycle to cycle \citep{1988MmSAI..59..147P}.
The time scale of the dips in the RB is similar to that 
shown by some dipping sources \citep[e.g.][]{1984PASJ...36..731K},
and, as we said, the fact that the dips go twice below the pre-burst flux
indicates that, as in dipping sources,
the persistent emission is also partially obscured.
Most importantly, the RB dips spectrum is consistent
with a larger portion of the flow
coming in the line of sight.
Unlike in all other sources, however, in the RB
the dips only occur during bursts.
Likely this means that the inclination angle is lower,
since at a high inclination angle dips in the persistent emission
should also be visible.

None the less, obscuration phenomena of the surface
might be possible even at low inclinations.
3D MHD simulations by
\citet{2008ApJ...673L.171R,2012MNRAS.421...63R} and \citet{2008MNRAS.386..673K}
showed that two different regimes of accretion
on magnetized stars like AMXPs are possible
in presence of a tilt ($\lesssim 30\degr$) between
the rotational axis and the magnetic dipole.
At large $\dot{M}$, in what they call the ``boundary layer'' regime,
matter can accrete directly through the magnetosphere
via narrow and tall ``tongues'' due to Raleigh-Taylor instability,
forming a belt-shaped hot region on the surface of the star.
Instead at lower $\dot{M}$, in the so-called ``magnetospheric regime'',
the accretion flow forms few (two or four) ordered funnel streams toward the magnetic poles.

If these predictions are correct, for a low inclination
the appearance of obscuring funnels
is indeed to be expected as $\dot{M}$ goes down.
Furthermore, as these funnels are nothing but transient accretion episodes at low $\dot{M}$,
they might help explain the onset of the type II burst activity
at the transition from the soft to the hard state.
However, if this change in geometry is behind the obscuration,
it remains unclear why double peaks would not be
an overall characteristic of all bursts in the low $\dot{M}$ state,
and also whether the a burst-funnel interaction
could reproduce the observed dip durations.

\subsection{Nature of 0.25 Hz QPO}

QPOs in LMXBs, no matter how difficult to explain, are always ascribed
to a phenomenon in the accretion flow \citep[see][and references
  therein]{2006csxs.book...39V}. The only exceptions are mHz QPOs
which are thought to be due to marginal nuclear burning on the NS surface
(\citealt{2001A&A...372..138R}; \citealt*{2007ApJ...665.1311H}).
However, those QPOs are interrupted by type I X-ray bursts \citep{2008ApJ...673L..35A}
and have frequencies one to two orders of magnitude smaller than the 0.25 Hz QPOs in the
RB. Thus, it is most likely that the dipping behavior and the 0.25 Hz
QPO in the RB are related to the accretion flow.

We explore in this section two possible scenarios to explain the QPO:
periodic obscurations
and modulation of the accretion flow via
interactions between the disc and the magnetosphere.

\subsubsection{QPOs in dipping sources}

At least three dipping bursting LMXBs (XB 1323-619,
\citealt{1999ApJ...511L..41J}; EXO 0748-676,
\citealt{1999ApJ...516L..91H}; and 4U 1746-37,
\citealt{2000ApJ...531..453J})  show $\sim$1~Hz QPOs.
Their rms amplitudes are about 10 per cent, and
the energy dependence of the rms amplitude is nearly flat.  The
fractional rms amplitude was found to be consistent with being
constant during the persistent emission, the X-ray dips and, most
importantly, type I X-ray bursts.
The fact that the $\sim$1~Hz QPOs persist during a dip
with unchanged rms indicates that they are formed in a region
at a smaller radius than that where the obscuration takes place.

It seems therefore likely that the $\sim$1~Hz QPOs detected in dipping LMXBs
are caused by matter periodically obscuring our line of sight to the NS,
perhaps due to Lense-Thirring precession of
a tilted inner accretion disk around a compact object \citep{2012ApJ...760L..30H}.
Despite the rough match in frequency and amplitude,
no dips have ever been observed in the persistent emission of the RB
outside a burst, which
seems at odds with a system viewed at a high inclination.
Also, the $\sim$1~Hz QPOs of dipping sources
span a much larger range of accretion states,
whereas the QPO in the RB appears to be confined
to the transition between the island and banana states.
Therefore we consider this scenario not likely to explain the 0.25~Hz QPO in the RB.

\subsubsection{QPOs from disc-magnetosphere interactions}

First reported by \citet{2000IAUC.7358....3V},
the 1~Hz QPO that is observed in the final stages of the outbursts of SAX J1808.4-3658
was extensively studied by \citet{2009ApJ...707.1296P}.
They report a QPO frequency in the (0.8-1.6)~Hz range,
and an rms amplitude that grows monotonically with energy up to about 17 keV.
A 1~Hz QPO of similar properties has been reported from NGC 6440 X-2
by \citet{2013ApJ...771...94P}, who suggest this feature has the same origin
in both these AMXPs.

The most likely origin for the QPO is an accretion flow instability arising 
at the inner edge of the disk \citep{2009ApJ...707.1296P}.
Among the variety of mechanisms at hand,
the most promising seems to be the one first proposed by
\citet{1993ApJ...402..593S}
and further explored
by \citet{2010MNRAS.406.1208D,2012MNRAS.420..416D}.  In this model an
instability can develop when the magnetosphere truncates the inner
accretion disc at a radius $r\sub{in}$ that is outside but still close
to the corotation radius $r\sub{cor}$.  A new disk solution known as
``dead disc'' is then found 
\citep{1977SvAL....3..138S,2010MNRAS.406.1208D,2012MNRAS.420..416D}
which is alternative to the ``propeller'' regime \citep{1975A&A....39..185I},
where the transfer of angular momentum effectively expels matter from
the system, and the inner disc radius keeps increasing as $\dot{M}$ decreases.
When $r\sub{in}\gtrsim r\sub{cor}$ the transfer of angular
momentum happens from the neutron star to the disk, spinning down the
neutron star and changing the radial gas density distribution at the
inner disc region without matter expulsion.  In a dead disc
$r\sub{in}$ does not evolve far from $r\sub{cor}$ and the matter keeps
flowing in from the outer disc regions piling up there. If certain
conditions are met at the magnetosphere-disk interaction
\citep{2010MNRAS.406.1208D,2012MNRAS.420..416D} a critical pressure
is eventually reached under which the gas slips inside $r\sub{cor}$,
thereby initiating a sudden accretion event. After that, the
accumulated mass reservoir is depleted, and a new cycle begins.

This instability is a potentially interesting model
for the 0.25~Hz QPO in the RB.
The frequency of the instability depends on the viscous timescale at the
inner disk edge, which in turn depends on the mass accretion rate and on 
the location of the inner disk (see e.g., Eq.~9 in \citealt{2013ApJ...771...94P}). 
Unfortunately, the inner disk radius is not known for the Rapid Burster,
since neither the spin frequency of the neutron star
nor the strength of its magnetic dipole are known.

Also, a slightly growing rms-$E$ relation is observed in the RB,
which is also observed in the other two sources.
While the increase is weaker than in SAX J1808.4-3658
or NGC 6440 X-2, this can be explained with the smaller fractional amplitude
of the QPO in the RB.
\citet{2013ApJ...771...94P} explain the monotonic increase in rms with energy
in SAX J1808.4-3658 and NGC 6440 X-2 arguing that
as the instability is caused by changes in $\dot{M}$,
and the latter change the density and temperature of the corona
where the photons are upscattered,
the QPO fluctuations will translate to a varying hardness of the spectrum,
making the high energy flux more variable than the low energy one.
In case of a strong persistent emission,
as is the case in the RB, the effect will obviously be weaker.

Differences are, however, clearly present.
The QPO in the two AMXPs
cannot be fit with a Lorentzian because of its fast decline
in power at low frequencies.
This is however not seen in the RB QPO.
Furthermore, the rms amplitude of the QPO
is more than one order of magnitude larger
in the two other sources than in the RB.
This is because nearly no accretion takes place during the QPO minima
in SAX J1808.4-3658 and NGC 6440 X-2,
which show the oscillation at moderately low luminosities.
The higher count rates and the different QPO shape observed in the RB 
are therefore challenging to this scenario.
Even though the results from \citet{2010MNRAS.406.1208D}
show that for certain ranges of parameters
accretion can in fact continue to take place
in the minima of the QPO,
the instability is predicted to produce a much larger modulation
of the mass accretion rate $\dot{M}$ than observed in the RB,
when at the time of the double-peaked bursts
$\dot{M} \simeq 4 \times 10^{-9} M_{\odot}\textrm{ yr}^{-1}$,
roughly one order of magnitude larger than what observed in the two other sources
at the time of the appearance of the QPO there.

If however we again invoke a two-channel accretion
(see Sec.~\ref{sec:multipeak}),
with a polar and an equatorial channels,
one controlled by the instability and the other one unaffected by it,
the lower amplitude would naturally follow.
For example, while the centrifugal barrier could act to modulate the equatorial accretion,
matter could be leaking from above and below it,
and accrete along the $B$ lines.

Also involving two-channel accretion, the disc-magnetospheric interactions described in
\citet{2008ApJ...673L.171R} and \citet{2008MNRAS.386..673K}
can as well explain the appearance of periodicities in the lightcurve,
via accretion-induced hotspots.
While at high $\dot{M}$ sporadically forming ``tongues'' hit the star in random places,
so that the light curve is irregular,
at low $\dot{M}$ the funnels driving matter from the disc to the magnetic poles
are relatively stable, so that the position of the hot spots is almost constant,
resulting in a modulation of the lightcurve.
Again, the problem with this model is that it does not explain
why the QPO is short-lived, rather than being an overall characteristic of
the island state.

	\subsection{Linking the dips and the QPO}

In Table~\ref{tab:mechanisms} we list the possible scenarios for
the dips and the QPO, and which relevant observables
they are capable of explaining.

The only model that is not ruled out by any observable involves
a relatively short-lived
accretion state consisting of two different accretion channels,
an equatorial and a polar one,
in a system viewed at low inclination.
As part of the inner accretion flow in this particular state
goes along the magnetic field lines to the NS poles,
it is then easier for material to come in the line of sight
if the inclination angle is low,
and for the burst to disturb the accretion flow,
thereby causing the flux to dip below the pre-burst level,
even if the burst flux is sub-Eddington.

Since the 0.25~Hz QPO disappears during some of the dips,
its place of origin would also seem to be obscured by the same material.
This suggests that the QPO originates close to the NS and closer
than the X-ray emitting part of the accretion disk.
The latter region probably reaches out to 10$^4$ km
\citep{1984PASJ...36..741M,1986ApJ...308..635M}, so the obscuring
material would have to be substantially closer than that.
If the QPO originates in the equatorial flow,
the accretion channel along the magnetic field could be disturbed
by the burst so that, given the low inclination, the radiation from the
NS surface and the equatorial flow would be obscured, making the
dips deeper than the pre-burst emission and causing the temporary
disappearance of the QPO.
If instead the QPO originates in the disrupted polar flow,
its disappearance at the time
when the dips are deeper than the pre-burst emission would still naturally follow.

To match the observations,
this peculiar configuration should be short-lived,
appearing roughly at the same time when the type II burst do,
and disappearing within days.
It is generally thought that in the banana state,
at larger mass-accretion rates, a geometrically thin and optically thick disc
probably extends down to very close to the NS surface,
whereas the lack of dips in the island state,
at lower mass-accretion rates,
could be due to a quasi-spherically symmetric accretion
via an optically thin corona
\citep*[see e.g. review by][]{2007A&ARv..15....1D}.
The rarity of the double-peaked bursts and of the 0.25~Hz QPO
would then match the fact that they occur in a particular accretion state of the RB,
when part of the accretion flow is starting
to become diverted to the type II bursting mechanism.

\section{Conclusions}

We have reported the coincident observation
of non-PRE, double-peaked type I bursts and a LF QPO with a frequency of about 0.25~Hz.
They occur at the transition from the banana to the island state,
when the geometry of the accretion flow is expected to change.
This transition also coincides with the appearance of type II bursts.
Despite being sometimes at roughly the same count rate as the pre-burst persistent emission,
the dip between the two peaks has a distinctively different spectrum,
with a much larger hydrogen column density required to fit the data.
This is suggestive of some kind of temporary obscuration of the NS surface.
Also, the dips in two bursts go below the pre-burst emission level,
meaning part of the accretion flow is also obscured, or perhaps even disrupted.
No dips are ever observed in the persistent emission outside a burst.

We have explored several models to explain the dips, the QPO,
and the peculiar accretion state in which they appear,
including confining mechanisms on the NS surface,
obscuration of the surface and inner accretion flow in a high-inclination system,
and accretion instabilities borne out of the interaction between the accretion disc and the magnetosphere.

Although no single model can explain all the observables,
we favour a two-channel accretion state,
transitional between the soft and hard state.
Mass accretion onto the magnetic poles
would allow for the burst to disturb the accretion flow
even if the burst flux is sub-Eddington.
If the source is viewed at a low inclination,
the temporary disruption of the polar accretion channel by the burst
could then obscure the NS surface,
which would cause the flux to go below its pre-burst level,
and in some cases even the inner equatorial accretion flow.
If the instability causing the QPO
controlled the equatorial channel,
the disappearance of the QPO in the deepest dips
could also be explained.

Further investigation is necessary to develop this model
beyond this qualitative picture.
It remains to be seen whether such an accretion geometry could indeed
be sustainable, confined to a very narrow range of accretion rates,
and whether this scenario could reproduce the fairly long dips.
The difficulty of constraining such a model notwithstanding,
the obscuration that is responsible from the dip
must be connected with the disc instability
that is responsible for the QPO and the type II bursts.

\section*{Acknowledgments}

We wish to thank
Diego Altamirano,
Yuri Cavecchi,
Jeroen Homan,
Daniela Huppenkothen
and Hauke Worpel
for useful discussions
and occasional guidance offered during the preparation of this paper.
TB and AW are members of
an International Team in Space Science on type I X-ray bursts
sponsored by the International Space Science Institute (ISSI) in
Bern, and we thank ISSI for hospitality during part of this work.

\footnotesize{
  \bibliographystyle{mn2e}
  \bibliography{paper}
}

\label{lastpage}
\end{document}